\documentclass{article}
\usepackage{amsmath,graphicx}
\usepackage{booktabs} 
\usepackage{multirow}
\usepackage{amsfonts}
\usepackage{hyperref}
\usepackage[preprint]{spconf}

\copyrightnotice{\copyright\ IEEE 2023}
\toappear{To appear in {\it Proc.\ ICASSP2023}}

\title{Self-Supervised Learning-Based Source Separation for Meeting Data}
\name{Yuang Li, Xianrui Zheng\sthanks{*Zheng is supported by an Amazon Studentship.}, Philip C. Woodland~\thanks{This work has been performed using resources provided by the Cambridge Tier-2 system operated by the University of Cambridge Research Computing Service (\url{www.hpc.cam.ac.uk}) funded by EPSRC Tier-2 capital grant EP/T022159/1.}}

\address{Cambridge University Engineering Dept., Trumpington St., Cambridge, CB2 1PZ U.K.\\
        \texttt{\small\{yl807, xz396, pcw\}@eng.cam.ac.uk}}
\begin{document}
\ninept

\maketitle

\begin{abstract}

Source separation can improve automatic speech recognition (ASR) under multi-party meeting scenarios by extracting single-speaker signals from overlapped speech. Despite the success of self-supervised learning models in single-channel source separation, most studies have focused on simulated setups. In this paper, seven SSL models were compared on both simulated and real-world corpora. Then, we propose to integrate the best-performing model WavLM into an automatic transcription system through a novel iterative source selection method. To improve real-world performance, time-domain unsupervised mixture invariant training was adapted to the time-frequency domain. Experiments showed that in the transcription system when source separation was inserted before an ASR model fine-tuned on separated speech, absolute reductions of 1.9\% and 1.5\% in concatenated minimum-permutation word error rate for an unknown number of speakers (cpWER-us) were observed on the AMI dev and test sets.

\end{abstract}

\section{Introduction}
\label{sec:intro}

In everyday conversation, speech overlap occurs frequently, leading to the degradation of speech intelligibility and a significant challenge for ASR systems. To overcome this problem, source separation can be applied to estimate single-speaker signals. Early studies like deep clustering~\cite{hershey2016deep} and utterance-level permutation invariant training~\cite{kolbaek2017multitalker} treated single-channel source separation in the time-frequency (T-F) domain and ignored phase information, so the task was to estimate single-speaker magnitude spectrum. Recently time-domain methods have become popular owing to the ability to jointly model between magnitudes and phases. Starting from TasNet~\cite{luo2018tasnet} which directly predicts waveforms with an encoder-decoder structure, various time-domain architectures were proposed~\cite{luo2019conv, luo2020dual}. So far, Transformer-based models~\cite{subakan2021attention, jingjing2020dual} demonstrated the most promising performance.\\
\indent Supervised training requires single-speaker signals as targets, but obtaining such references from real overlapped signals is challenging, so a common approach is to combine multiple single-speaker segments~\cite{hershey2016deep}. To improve realism, noises and reverberations were added~\cite{wichern2019wham, maciejewski2020whamr}, but the mismatch between real and synthetic data is still problematic. To utilise real overlapped speech, researchers have proposed unsupervised training methods which use pseudo-labels from well-trained separation models~\cite{han2022heterogeneous} or exploit the consistency between separated sources and the mixture~\cite{wisdom2020unsupervised}. Sivaraman et al.\cite{sivaraman2022adapting} showed that unsupervised mixture invariant training (MixIT) was effective on AMI but ASR-based evaluation was not included and they focused on time-domain rather than T-F domain masking-based models.\\
\indent Self-supervised learning (SSL) benefits various speech-related downstream tasks by learning from a large amount of unlabelled signals~\cite{yang2021superb}. With a Conformer network~\cite{gulati2020conformer} as the downstream model, WavLM~\cite{chen2021wavlm} set the state-of-the-art separation performance on LibriCSS~\cite{chen2020continuous}. Huang et al.\cite{huang2022investigating} further compared 13 SSL models on LibriMix~\cite{cosentino2020librimix} to demonstrate their effectiveness. However, these comparisons were limited to simulated datasets. Furthermore, SSL models were frozen and only used to generate acoustic representations.\\
\indent In this paper, the goal is to perform a thorough analysis of SSL models on real-world source separation tasks. For the traditional utterance-wise separation task, seven SSL models were compared on a reverberated version of LibriMix~\cite{cosentino2020librimix} (R-LibriMix), LibriCSS~\cite{chen2020continuous}, and AMI~\cite{kraaij2005ami} by using both signal-based and ASR-based criteria. The potential of SSL models was fully exploited through a two-phase fine-tuning schedule where the lightweight downstream model was first trained and then the SSL model was fine-tuned. Experiments were then extended to a real-world application: a meeting transcription system. To alleviate domain mismatch, the T-F domain unsupervised MixIT using phase-sensitive masking (PSM) was introduced to fine-tune the model with real overlapped data. The separation model has several output sources and the desired one needs to be chosen automatically based on speaker information. To this end, the proposed source selection method computes and refines average speaker embeddings iteratively according to speaker labels provided by diarisation.\\
\indent In the rest of this paper, sec.~\ref{sec:ssl} reviews SSL models. Sec.~\ref{sec:method} presents source separation framework, supervised and unsupervised training methods, and the proposed iterative source selection method. Sec.~\ref{sec:setup} and sec.~\ref{sec:results} provides experimental setups and results. Sec.~\ref{sec:conclusions} concludes this paper. Audio samples are publicly available~\footnote{\url{https://sites.google.com/view/ssl-ss/home}}.

\section{SSL Models for Speech Signals}
\label{sec:ssl}

SSL models provide informative representations for downstream tasks and good initialisation for further fine-tuning. These models can be divided into two categories: generative models which reconstruct raw features and discriminative models which estimate discretised speech representations generated by clustering or quantisation. \textbf{TERA}~\cite{liu2021tera} is a self-supervised Transformer encoder trained on a generative task where input spectrograms were randomly masked during training. The model was optimised using the L1 loss to reconstruct the original spectrograms. \textbf{Wav2Vec2}~\cite{baevski2020wav2vec} used raw waveforms as input and masked features in the latent space. It introduced a quantisation module that discretised latent features as training targets. The model was trained through a contrastive learning task where it should discriminate the true quantised features at masked time steps from a set of distractors. \textbf{HuBert}~\cite{hsu2021hubert} adopted an offline clustering approach to generate discrete labels. The model estimated the distribution of these pseudo-labels and then the cross-entropy loss was computed over the masked region. \textbf{UniSpeech-SAT} and \textbf{WavLM}~\cite{chen2022unispeech, chen2021wavlm} are variants of HuBert proposed to facilitate the extraction of speaker identity. A key feature shared by them is the use of utterance mixing augmentation, which involves combining the main utterance with a randomly selected segment from a different speaker.\\

\section{Methodology}
\label{sec:method}

\subsection{SSL Representations for Source Separation}

\begin{figure}[ht]

\centering
  \includegraphics[width=0.9\linewidth]{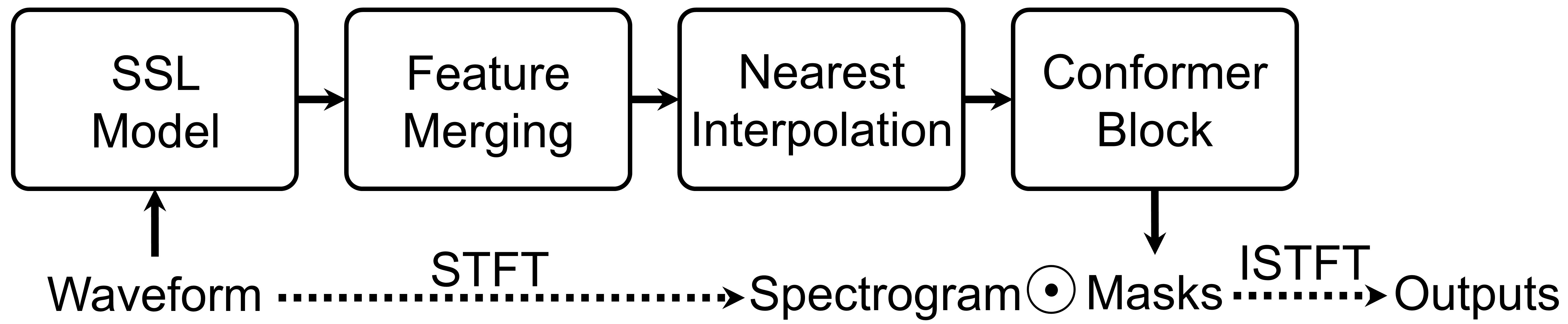}
  \caption{Framework of T-F domain source separation using SSL representations.}
  \label{fig:ss}
\end{figure}

\indent The SSL-based source separation framework is shown in Figure~\ref{fig:ss}. The SSL model processes time domain waveforms and generates SSL representations. Then, to exploit multi-level information, features from every SSL Transformer layer are merged using a weighted sum~\cite{huang2022investigating}. Following feature merging, the nearest interpolation stage increases the temporal resolution of SSL representations so they have the same temporal resolution as spectrograms. Finally, a single Conformer block estimates multiple T-F masks, which are used to recover the separated signals. The Conformer block has significantly fewer parameters than the downstream models in previous works~\cite{chen2021wavlm, huang2022investigating} so the SSL model plays the key role in source separation. Note that the time-domain model was not used as SSL features have a similar stride of spectrograms whereas the time-domain model tends to use a much smaller stride.

\indent To optimise the separation model, the standard approach uses supervised permutation invariant training (PIT)~\cite{kolbaek2017multitalker} which selects the minimum loss of all possible permutations. In this paper, when the model was trained on synthetic datasets, the loss was computed following Eqn.(~\ref{eq:pit_loss}) where a phase-sensitive target~\cite{erdogan2015phase} spectrogram was used.

\begin{equation}
\setlength{\abovedisplayskip}{2pt}
\setlength{\belowdisplayskip}{2pt}
\label{eq:pit_loss}
\mathcal{L}_{\text{PIT}} = \min_{\phi \in P}\sum_{(i, j)\in\phi}||\textbf{\textit{M}}_{i}\odot|\textbf{\textit{Y}}| - |\textbf{\textit{X}}_j|cos(\theta_Y - \theta_{X_j})||_{F}^{2}
\end{equation}

\noindent where $\phi$ denotes for a permutation; $P$ is the set of all possible permutations; $\textbf{\textit{Y}}$ is the T-F domain mixture; $\textbf{\textit{M}}_i$ is a output mask; $\textbf{\textit{X}}_j$ is a target spectrogram; $\theta_Y$ and $\theta_{X_j}$ are the phases of $\textbf{\textit{Y}}$ and $\textbf{\textit{X}}_j$; $||\cdot||_F$ is the Frobenius norm.

\subsection{T-F domain Mixture Invariant Training (MixIT)}

PIT only works with synthetic data whereas MixIT can use real mixtures as labels by assuming the output sources can be remixed into these labels. MixIT first creates the mixture of mixtures (MoMs) by combining real overlapped signals. Then, the model takes MoMs as input and estimates several audio streams. To compute the loss, MixIT exhaustively searches for the remix with the minimum error.\\
\indent The original MixIT procedure was used for time-domain source separation, in which case outputs can be remixed by adding samples of waveforms, but in the T-F domain, it is inappropriate to directly add magnitudes because of the phase difference. Therefore, MixIT was adapted to the T-F domain with PSM because the time domain remix is equivalent to the remix of phase-sensitive targets so the loss function becomes Eqn.(~\ref{eq:mit_loss}) which has a similar form to PIT. The difference is that the prediction is a mixture ($\sum_{i\in\mathbf{I}}\textbf{\textit{M}}_{i}\odot|\textbf{\textit{Y}}|$) instead of a single spectrum ($\textbf{\textit{M}}_{i}\odot|\textbf{\textit{Y}}|$).\\

\begin{equation}
\setlength{\abovedisplayskip}{0pt}
\setlength{\belowdisplayskip}{2pt}
\label{eq:mit_loss}
\mathcal{L}_{\text{MixIT}} = \min_{\phi \in \mathcal{M}}\sum_{(\mathbf{I}, j)\in\phi}||(\sum_{i\in\mathbf{I}}\textbf{\textit{M}}_{i}\odot|\textbf{\textit{Y}}|) - |\textbf{\textit{X}}_j|cos(\theta_Y - \theta_{X_j})||_{F}^{2}
\end{equation}

\noindent where $\phi$ is a remix; $\mathcal{M}$ denotes all possible remixes; $\textbf{I}$ is a set of output streams that should be combined; $\textbf{\textit{Y}}$ is a T-F domain MoM; $\textbf{\textit{M}}_i$ is a output mask; $\textbf{\textit{X}}_j$ is a target spectrogram; $\theta_Y$ and $\theta_{X_j}$ are the phases of $\textbf{\textit{Y}}$ and $\textbf{\textit{X}}_j$.\\
\indent MixIT does not explicitly guide the model to separate a single speaker’s signal into one output stream which may lead to over-separation, so semi-supervised learning which combines PIT
with MixIT is more effective.

\subsection{Iterative Source Selection}
Compared to source separation without a target speaker, speaker extraction is a more realistic task where the model estimates the target speaker’s speech from the mixture. To identify the target speaker, the speaker embedding extracted from enrolment audio by an auxiliary branch is commonly injected into the main network~\cite{delcroix2018single}. In this paper, speaker extraction was decoupled into source separation and source selection. For source selection, it was assumed that utterances come from a diarisation system that provides speaker labels. Consequently, an average speaker embedding can be derived from utterances of the same speaker and later used to select from output sources. However, these utterances are polluted by overlaped data, so speaker embeddings are iteratively refined and some outliers are removed. The detailed steps are as follows:
\begin{itemize}
  \setlength{\itemsep}{0pt}
  \setlength{\itemindent}{-1.5em}
    \item \textbf{Step 1}: Compute speaker embedding of every utterance.
    \item \textbf{Step 2}: Compute average embedding for each speaker.
    \item \textbf{Step 3}: For each speaker, remove part of outliers with high Euclidean distances to the average embedding.
    \item \textbf{Step 4}: Compute average speaker embedding without outliers.
    \item \textbf{Step 5}: Select output source whose embedding has the highest cosine similarity with the average speaker embedding.
    \item \textbf{Step 6}: Compute embedding of selected output sources.
    \item \textbf{Step 7}: Return to step 2.
\end{itemize}
\indent With source selection, separation models can be applied directly for speaker extraction without modifying the architecture and retraining. Moreover, our approach does not require enrollment audio and can be easily plugged into a transcription system.

\section{Experimental Setup}
\label{sec:setup}

\subsection{Datasets and Configurations}
For PIT experiments, LibriMix~\cite{cosentino2020librimix} was extended to R-LibriMix in this paper by simulating reverberations through the image method~\cite{allen1979image} and adding isotropic noise~\cite{habets2007generating}. The reverberated signals were used as targets, so the model focused on separation. For MixIT experiments, the beam-formed audio recorded by the distant microphone from AMI~\cite{kraaij2005ami} was used. MoMs were created with signal-to-noise ratio (SNR) between -5dB and 5dB. For evaluation, to enable both in-domain signal-based and real-world ASR-based comparisons, the synthetic R-LibriMix, the simulated LibriCSS~\cite{chen2020continuous} and the real AMI were used.\\
\indent The downstream model was a single Conformer block~\cite{chen2021continuous} with an attention dimension of 256, 4 attention heads, and 1024-dimensional hidden states. The number of output mask was 2 for supervised training (PIT) and 4 for semi-supervised training (PIT\&MixIT). Softmax replaced the Sigmoid activation for semi-supervised training as it introduced connections between masks which is consistent with the assumption of MixIT, contributing to faster and stable convergence. The front-end short-time Fourier transformation (STFT) used a 400-point Hann window with a stride of 160 points. The fine-tuning schedule had two phases. In the first phase, only the Conformer block was updated for 100k steps. In the second phase, the SSL model was unfrozen and fine-tuned for another 80k steps. For semi-supervised training, MixIT was used with a probability of 80\%, and batches were sampled from AMI. Otherwise, PIT was used with batches sampled from R-LibriMix.

\subsection{Transcription System}

\begin{figure}[ht]
\centering
  \includegraphics[width=0.9\linewidth]{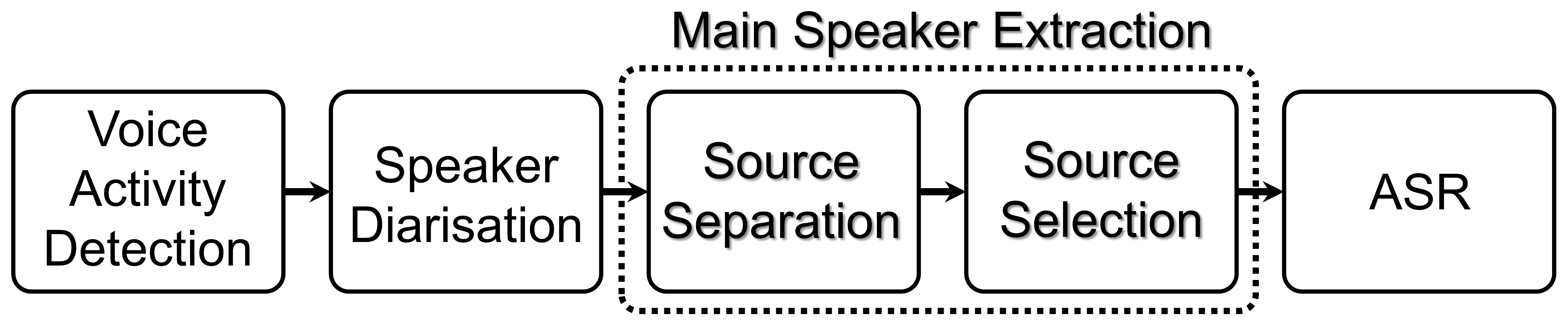}
  \caption{Flowchart of the automatic transcription system.}
  \label{fig:meeting}
\end{figure}

Figure~\ref{fig:meeting} is a flowchart for the transcription system. First, continuous recordings were cropped into utterances and tagged with speaker labels by voice activity detection (VAD) and diarisation. Then, overlaps were removed and the main speaker's speech signals were kept through source separation and selection. Finally, an ASR model was applied to produce transcriptions. The details are as follows: \textbf{1) VAD}: Wav2Vec2 was fine-tuned on AMI by adding a linear layer followed by the Softmax activation to give the probabilities of the presence/absence of speech components. The time interval between speech regions was constrained to be more than 0.04s, the same as the pipeline in~\cite{zheng2022tandem}. \textbf{2) Diarisation}: ECAPA-TDNN~\cite{dawalatabad2021ecapa} generated speaker embeddings with a window size of 3 seconds and stride of 1.5 seconds. Then, spectral clustering was applied with a maximum speaker number of 10. \textbf{3) Source Separation}: the WavLM model fine-tuned on R-LibirMix and AMI through semi-supervised training was chosen. \textbf{4) Source Selection}: Iterative source selection was used with speaker embeddings generated from ECAPA-TDNN. On the AMI dev set, it was found that WER notably decreased after two iterations while remaining roughly unchanged afterward. Hence the number of iterations was set to two for efficiency. Additionally, 60\% of outliers were removed to stabilise convergence.  \textbf{5) ASR}: Three ASR models that used Wav2Vec2-Robust~\cite{hsu2021robust} as the base model were compared: W2V2-SWB was fine-tuned on Switchboard~\cite{godfrey1992switchboard}; W2V2-AMI was fine-tuned on AMI; W2V2-AMI-Sep was a further fine-tuned version of W2V2-AMI, using separated audio from AMI.

\subsection{Evaluations}
For initial comparisons between SSL models, scale invariant SNR (SI-SNR)~\cite{luo2018tasnet} and word error rate (WER) was used. R-LibriMix provides reference signals, so the highest SI-SNRs of all possible permutations were computed. For LibriCSS~\footnote{ASR model for LibriCSS: \url{https://huggingface.co/facebook/wav2vec2-large-960h-lv60-self}} and AMI~\footnote{ASR model for AMI: \url{https://huggingface.co/facebook/wav2vec2-large-robust-ft-swbd-300h}}, only the transcriptions of utterances are available, so the lowest possible WERs using separated sources were found.\\
\indent Source selection was evaluated by comparing it to the oracle selection~\footnote{The oracle selection means choosing the source whose transcription best matches the ground truth} in terms of WERs. The selection accuracy was also calculated which was defined as the percentage of duration that the estimated selection matches the oracle selection.\\
\indent For the transcription system, traditional WER is difficult to compute as segments provided by the VAD can be very different from the start and end times of utterance in the reference. Therefore, the cpWER-us~\cite{zheng2022tandem} at the session level was used. First, the hypotheses and references were concatenated respectively for speakers in each session in chronological order. Then, WERs of all possible speaker permutations were computed and redundant speakers were removed. Finally, the lowest WER among them was chosen as the cpWER-us.

\section{Experimental Results}
\label{sec:results}

\subsection{Comparisons between SSL models}

As shown in Table~\ref{table:results_ssl}, when only the downstream model was trained, all SSL models outperformed the spectrogram on R-LibriMix and LibriCSS. However, on AMI dev and test sets, the spectrogram had lower WERs of 38.1\% and 40.5\% compared to Wav2Vec2 and HuBert whose WERs are 38.7\% and 38.3\% on AMI dev set and both 41.0\% on AMI test set. TERA was slightly worse than Wav2Vec2 and HuBert on R-LibriMix but performed better on AMI dev and test sets with WERs of 37.8\% and 40.3\% possibly because the spectral alternations used to pre-train TERA improved its generalisation. These comparisons showed that although R-LibriMix and LibriCSS were created to simulate meeting recordings, the results can be still inconsistent with a real corpus because of different noise types, reverberation, and speech content. If the SSL model is only trained on clean recordings of read texts, it can generalise poorly on real overlapped conversations. UniSpeech-SAT and WavLM achieved significantly better performance across all the datasets than other models even without an enlarged pre-training dataset. On the AMI corpus, UniSpeech-SAT achieved dev and test set WERs of 36.6\% and 39.6\% while WavLM had slightly higher WERs of 36.8\% and 40.1\%. It can be concluded that utterance mixing which allowed the model to be exposed to multiple speakers at the same time is the key to extracting robust representations for source separation. \\
\indent After the SSL models were fine-tuned, consistent performance improvements were observed as the trainable parameters increased significantly from 1.8M to 92M for all SSL models except for TERA. TERA has the fewest parameters of 23M so the improvements were the smallest. Through the fine-tuning experiments, it can be concluded that unfreezing SSL models is beneficial as it allows the model to learn task-specific representations and it will not lead to overfitting since cross-dataset performance on AMI was also improved. Although all SSL models were optimised for source separation directly, UniSpeech-SAT and WavLM were still notably better than Wav2Vec2 and HuBert which implied that an initialisation from a SSL task is crucial even for a task like source separation where large synthetic datasets can be automatically generated. Furthermore, if WavLM+ is compared to UniSpeech-SAT+, it can be seen that WavLM+ gained more benefit from a large pre-training dataset, as the performance gap between WavLM and WavLM+ was larger in most cases. For example, on AMI dev and test sets, the absolute WER improvements from WavLM to WavLM+ are 0.9\% and 0.6\%. However, for UniSpeech-SAT, only 0.3\% of absolute improvement was observed on AMI test set. In general, WavLM+ is the best-performing model, so it was used in the rest of the paper.

\begin{table}[t]
\setlength{\tabcolsep}{5.5pt}
\small
\centering
\begin{tabular}{c | c | c | c} 
 \toprule
  & R-LibriMix & LibriCSS & AMI\\
 Model & SI-SNR$\uparrow$ & WER$\downarrow$ & WER$\downarrow$\\
\midrule
No Separation & -0.43 / -0.44  & 14.7 & 41.2 / 43.8\\
Spectrogram & 3.15 / 2.96  & 12.4 & 38.1 / 40.5\\
 \midrule
 \multicolumn{4}{c}{Phase 1: train the downstream model.}\\
 \midrule
TERA~\cite{liu2021tera} & 4.80 / 4.60 & 11.2 & 37.8 / 40.3\\
Wav2Vec2~\cite{baevski2020wav2vec} & 5.31 / 5.18 & 11.4 & 38.7 / 41.0\\
HuBert~\cite{hsu2021hubert} & 5.12 / 4.93 & 11.4 & 38.3 / 41.0 \\
UniSpeech-SAT~\cite{chen2022unispeech} & 6.89 / 6.83 & 8.0 & 36.6 / 39.6\\
UniSpeech-SAT+~\cite{chen2022unispeech} & 7.01 / 6.87 & 7.9 & 36.4 / 39.3\\
WavLM~\cite{chen2021wavlm} & 6.51 / 6.36 & 7.7 & 36.8 / 40.1\\
WavLM+~\cite{chen2021wavlm} & \textbf{7.40} / \textbf{7.27} & \textbf{7.4} & \textbf{36.2} / \textbf{38.9}\\
\midrule
\multicolumn{4}{c}{Phase 2: fine-tune the base model.}\\
\midrule
TERA~\cite{liu2021tera} & 5.86 / 5.67 & 10.5 & 37.6 / 40.3\\
Wav2Vec2~\cite{baevski2020wav2vec} & 6.74 / 6.59 & 9.8 & 37.3 / 39.7\\
HuBert~\cite{hsu2021hubert} & 6.67 / 6.50 & 9.8 & 37.2 / 39.9 \\
UniSpeech-SAT~\cite{chen2022unispeech} & 7.73 / 7.65 & 7.1 & 35.5 / 38.6\\
UniSpeech-SAT+~\cite{chen2022unispeech} & 7.83 / 7.76  & \textbf{6.7} & 35.5 / 38.3\\
WavLM~\cite{chen2021wavlm} & 7.80 / 7.69 & 6.8 & 35.6 / 38.6 \\
WavLM+~\cite{chen2021wavlm} & \textbf{8.05} / \textbf{7.95} & \textbf{6.7} & \textbf{34.7} / \textbf{38.0}\\
\bottomrule
\end{tabular}
\caption{The comparisons between SSL models. For R-LibriMix and AMI, dev/test set SI-SNR or WERs are presented. For LibriCSS, the average WERs across all data subsets are provided. Models marked with "+" were pre-trained on a 94khr speech dataset. Other models were pre-trained on 960hr LibriSpeech dataset.}
\label{table:results_ssl}
\end{table}

\subsection{Results for MixIT and Source Selection}

Table~\ref{table:result_select} compares PIT with semi-supervised PIT\&MixIT. With the oracle source, PIT\&MixIT improved WERs with absolute reductions of around 1\% on AMI dev and test sets. The reason is that MixIT utilised real overlapped utterances, enriching the training data. Furthermore, MixIT is not limited to speech components. With four output sources, non-speech components like coughing and laughter were sometimes separated from speech signals. However, with iterative source selection, MixIT became less effective since more output sources increase the difficulty of source selection. For PIT\&MixIT, the selection accuracy was around 85\% but for PIT with only two sources, the selection accuracy was higher than 90\%. Therefore, in a real-world scenario, the conventional permutation invariant criteria can not be used to evaluate models with different numbers of output sources. The trade-off between separation performance and selection accuracy is more important. In this sense, semi-supervised training is still slightly better with lower WERs.

\begin{table}[t]
\centering
\small
\begin{tabular}{ c| c | c c} 
 \toprule
 & Oracle & \multicolumn{2}{c}{Iterative Selection}\\
 Training & WER$\downarrow$ & WER$\downarrow$ & Selection Acc.$\uparrow$\\ 
 \midrule
 PIT & 34.7 / 38.0 & 36.3 / 39.4 & 93.3 / 92.5\\
 PIT\&MixIT & 33.6 / 37.0 & 35.5 / 39.1 & 87.0 / 85.4\\
 \bottomrule
\end{tabular}
\caption{The comparisons of PIT and semi-supervised PIT\&MixIT on AMI. Dev/Test set WERs and selection accuracy are provided.}
\label{table:result_select}
\end{table}

\subsection{Results for the Transcription System}

Table~\ref{table:result_transcription} shows the influence of source separation on the transcription system. For W2V2-SWB, an ASR model trained on non-overlapped speech, without source separation, the cpWER-us was 43.9\% and 46.0\% on AMI dev and test sets. After source separation was incorporated, cpWER-us decreased to 40.8\% and 43.8\%. However, if the ASR model was fine-tuned on AMI (W2V2-AMI), source separation harmed the performance and a large increase of deletion errors was observed. A possible reason is that W2V2-AMI has seen overlapped data, so it learned to ignore the speaker in the background. After separation, more speech components may be incorrectly recognised as background. Moreover, the separation model introduced out-of-domain data like utterances with long silences and system noise. To tackle these issues, the W2V2-AMI model was fine-tuned on separated data resulting in W2V2-AMI-Sep which contributed to the lowest cpWER-us with absolute reductions of 1.9\% and 1.5\% on AMI dev and test sets compared to the system without source separation using W2V2-AMI. W2V2-AMI-Sep also displayed better performance using non-separated data with absolute cpWER-us reductions of 0.7\% and 0.7\% on AMI dev and test sets compared to W2V2-AMI which implied that overlaps can harm the training of ASR models.

\begin{table}[t]
\centering
\small
\begin{tabular}{ c | c | c } 
 \toprule
ASR & Separation & cpWER-us$\downarrow$\\
\midrule
W2V2-SWB & $\times$ & 43.9 / 46.0\\
W2V2-SWB & \checkmark & 40.8 / 43.8\\
W2V2-AMI & $\times$ & 34.2 / 35.1\\
W2V2-AMI & \checkmark & 36.4 / 36.8\\
W2V2-AMI-Sep & $\times$ & 33.5 / 34.4\\
W2V2-AMI-Sep & \checkmark & \textbf{32.3} / \textbf{33.6}\\
 \bottomrule
\end{tabular}
\caption{The comparisons of transcription systems with or without separation on AMI dataset. Dev/test set cpWER-us are provided.}
\label{table:result_transcription}
\end{table}

\section{Conclusions}
\label{sec:conclusions}

This paper investigated seven SSL models for source separation. We emphasised the importance of using real meeting data for evaluation as the performance may not be consistent with simulated datasets. Experiments showed that SSL models are effective not only as feature extractors but as trainable base models. However, SSL models trained on clean single-speaker signals such as Wav2Vec2 and HuBert have limited generalisability on real data. For the transcription system, source separation can be directly incorporated after diarisation and before ASR using the proposed iterative source selection method. To maximise the performance, MixIT was adopted to use real mixtures for training and the ASR model was further fine-tuned on separated audio. As the result, cpWER-us was notably reduced on AMI.




\clearpage
\bibliographystyle{IEEEbib}
\bibliography{refs.bib}

\begin{thebibliography}{10}

\bibitem{hershey2016deep}
J.~R. Hershey, Z.~Chen, J.~Le~Roux, and S.~Watanabe,
\newblock ``Deep clustering: Discriminative embeddings for segmentation and
  separation,''
\newblock {\em Proc. ICASSP}, 2016.

\bibitem{kolbaek2017multitalker}
M.~Kolb{\ae}k, D.~Yu, Z.~Tan, and J.~Jensen,
\newblock ``Multitalker speech separation with utterance-level permutation
  invariant training of deep recurrent neural networks,''
\newblock {\em IEEE Transactions on Audio, Speech, and Language Processing},
  vol. 25, no. 10, pp. 1901--1913, 2017.

\bibitem{luo2018tasnet}
Y.~Luo and N.~Mesgarani,
\newblock ``{TasNet}: time-domain audio separation network for real-time,
  single-channel speech separation,''
\newblock {\em Proc. ICASSP}, 2018.

\bibitem{luo2019conv}
Y.~Luo and N.~Mesgarani,
\newblock ``{Conv-TasNet}: Surpassing ideal time--frequency magnitude masking
  for speech separation,''
\newblock {\em IEEE Transactions on Audio, Speech, and Language Processing},
  vol. 27, no. 8, pp. 1256--1266, 2019.

\bibitem{luo2020dual}
Y.~Luo, Z.~Chen, and T.~Yoshioka,
\newblock ``{Dual-path RNN}: efficient long sequence modeling for time-domain
  single-channel speech separation,''
\newblock {\em Proc. ICASSP}, 2020.

\bibitem{subakan2021attention}
C.~Subakan, M.~Ravanelli, S.~Cornell, M.~Bronzi, and J.~Zhong,
\newblock ``Attention is all you need in speech separation,''
\newblock {\em Proc. ICASSP}, 2021.

\bibitem{jingjing2020dual}
J.~Chen, Q.~Mao, and D.~Liu.,
\newblock ``Dual-path {Transformer} network: Direct context-aware modeling for
  end-to-end monaural speech separation,''
\newblock {\em Proc. Interspeech}, 2020.

\bibitem{wichern2019wham}
G.~Wichern, J.~Antognini, M.~Flynn, L.~R. Zhu, E.~McQuinn, D.~Crow, E.~Manilow,
  and J.~L. Roux,
\newblock ``{WHAM!}: Extending speech separation to noisy environments,''
\newblock {\em Proc. Interspeech}, 2019.

\bibitem{maciejewski2020whamr}
M.~Maciejewski, G.~Wichern, E.~McQuinn, and J.~Le~Roux,
\newblock ``{WHAMR!}: Noisy and reverberant single-channel speech separation,''
\newblock {\em Proc. ICASSP}, 2020.

\bibitem{han2022heterogeneous}
J.~Han and Y.~Long,
\newblock ``Heterogeneous separation consistency training for adaptation of
  unsupervised speech separation,''
\newblock {\em arXiv preprint arXiv:2204.11032}, 2022.

\bibitem{wisdom2020unsupervised}
S.~Wisdom, E.~Tzinis, H.~Erdogan, R.~Weiss, K.~Wilson, and J.~Hershey,
\newblock ``Unsupervised sound separation using mixture invariant training,''
\newblock {\em Proc. NeurIPS}, 2020.

\bibitem{sivaraman2022adapting}
A.~Sivaraman, S.~Wisdom, H.~Erdogan, and J.~R. Hershey,
\newblock ``Adapting speech separation to real-world meetings using mixture
  invariant training,''
\newblock {\em Proc. ICASSP}, 2022.

\bibitem{yang2021superb}
S.~Yang, P.~Chi, Y.~Chuang, C.~J. Lai, K.~Lakhotia, Y.~Y Lin, A.~T Liu, J.~Shi,
  X.~Chang, G.~Lin, et~al.,
\newblock ``{SUPERB}: Speech processing universal performance benchmark,''
\newblock {\em Proc. Interspeech}, 2021.

\bibitem{gulati2020conformer}
A.~Gulati, J.~Qin, C.~Chiu, N.~Parmar, Y.~Zhang, J.~Yu, W.~Han, S.~Wang,
  Z.~Zhang, Y.~Wu, et~al.,
\newblock ``Conformer: Convolution-augmented transformer for speech
  recognition,''
\newblock {\em Proc. Interspeech}, 2020.

\bibitem{chen2021wavlm}
S.~Chen, C.~Wang, Z.~Chen, Y.~Wu, S.~Liu, Z.~Chen, J.~Li, N.~Kanda,
  T.~Yoshioka, X.~Xiao, et~al.,
\newblock ``{WavLM}: Large-scale self-supervised pre-training for full stack
  speech processing,''
\newblock {\em IEEE Journal of Selected Topics in Signal Processing}, 2022.

\bibitem{chen2020continuous}
Z.~Chen, T.~Yoshioka, L.~Lu, T.~Zhou, Z.~Meng, Y.~Luo, J.~Wu, X.~Xiao, and
  J.~Li,
\newblock ``Continuous speech separation: Dataset and analysis,''
\newblock {\em Proc. ICASSP}, 2020.

\bibitem{huang2022investigating}
Z.~Huang, S.~Watanabe, S.~Yang, P.~Garc{\'\i}a, and S.~Khudanpur,
\newblock ``Investigating self-supervised learning for speech enhancement and
  separation,''
\newblock {\em Proc. ICASSP}, 2022.

\bibitem{cosentino2020librimix}
J.~Cosentino, M.~Pariente, S.~Cornell, A.~Deleforge, and E.~Vincent,
\newblock ``{LibriMix}: An open-source dataset for generalizable speech
  separation,''
\newblock {\em arXiv preprint arXiv:2005.11262}, 2020.

\bibitem{kraaij2005ami}
W.~Kraaij, T.~Hain, M.~Lincoln, and W.~Post,
\newblock ``The {AMI} meeting corpus,''
\newblock {\em Proc. International Conference on Methods and Techniques in
  Behavioral Research}, 2005.

\bibitem{liu2021tera}
A.~T Liu, S.~Li, and H.~Lee,
\newblock ``{TERA}: Self-supervised learning of transformer encoder
  representation for speech,''
\newblock {\em IEEE Transactions on Audio, Speech, and Language Processing},
  vol. 29, pp. 2351--2366, 2021.

\bibitem{baevski2020wav2vec}
A.~Baevski, Y.~Zhou, A.~Mohamed, and M.~Auli,
\newblock ``{Wav2Vec} 2.0: A framework for self-supervised learning of speech
  representations,''
\newblock {\em Proc. NeurIPS}, 2020.

\bibitem{hsu2021hubert}
W.~Hsu, B.~Bolte, Y.~H. Tsai, K.~Lakhotia, R.~Salakhutdinov, and A.~Mohamed,
\newblock ``{HuBert}: Self-supervised speech representation learning by masked
  prediction of hidden units,''
\newblock {\em IEEE Transactions on Audio, Speech, and Language Processing},
  vol. 29, pp. 3451--3460, 2021.

\bibitem{chen2022unispeech}
S.~Chen, Y.~Wu, C.~Wang, Z.~Chen, Z.~Chen, S.~Liu, J.~Wu, Y.~Qian, F.~Wei,
  J.~Li, et~al.,
\newblock ``{UniSpeech-SAT}: Universal speech representation learning with
  speaker aware pre-training,''
\newblock {\em Proc. ICASSP}, 2022.

\bibitem{erdogan2015phase}
H.~Erdogan, J.~R Hershey, S.~Watanabe, and J.~Le~Roux,
\newblock ``Phase-sensitive and recognition-boosted speech separation using
  deep recurrent neural networks,''
\newblock {\em Proc. ICASSP}, 2015.

\bibitem{delcroix2018single}
M.~Delcroix, K.~Zmolikova, K.~Kinoshita, A.~Ogawa, and T.~Nakatani,
\newblock ``Single channel target speaker extraction and recognition with
  speaker beam,''
\newblock {\em Proc. ICASSP}, 2018.

\bibitem{allen1979image}
Jont~B Allen and David~A Berkley,
\newblock ``Image method for efficiently simulating small-room acoustics,''
\newblock {\em Journal of the Acoustical Society of America}, vol. 65, no. 4,
  pp. 943--950, 1979.

\bibitem{habets2007generating}
E.~AP Habets and S.~Gannot,
\newblock ``Generating sensor signals in isotropic noise fields,''
\newblock {\em Journal of the Acoustical Society of America}, vol. 122, no. 6,
  pp. 3464--3470, 2007.

\bibitem{chen2021continuous}
S.~Chen, Y.~Wu, Z.~Chen, J.~Wu, J.~Li, T.~Yoshioka, C.~Wang, S.~Liu, and
  M.~Zhou,
\newblock ``Continuous speech separation with {Conformer},''
\newblock {\em Proc. ICASSP}, 2021.

\bibitem{zheng2022tandem}
X.~Zheng, C.~Zhang, and P.~C Woodland,
\newblock ``Tandem multitask training of speaker diarisation and speech
  recognition for meeting transcription,''
\newblock {\em Proc. Interspeech}, 2022.

\bibitem{dawalatabad2021ecapa}
N.~Dawalatabad, M.~Ravanelli, F.~Grondin, J.~Thienpondt, B.~Desplanques, and
  H.~Na,
\newblock ``{ECAPA-TDNN} embeddings for speaker diarization,''
\newblock {\em Proc. Interspeech}, 2021.

\bibitem{hsu2021robust}
W.~Hsu, A.~Sriram, A.~Baevski, T.~Likhomanenko, Q.~Xu, V.~Pratap, J.~Kahn,
  A.~Lee, R.~Collobert, G.~Synnaeve, et~al.,
\newblock ``Robust {Wav2Vec} 2.0: Analyzing domain shift in self-supervised
  pre-training,''
\newblock {\em Proc. Interspeech}, 2021.

\bibitem{godfrey1992switchboard}
J.~J Godfrey, E.~C Holliman, and J.~McDaniel,
\newblock ``Switchboard: Telephone speech corpus for research and
  development,''
\newblock {\em Proc. ICASSP}, 1992.

\end{thebibliography}

\end{document}